\documentstyle[12pt,aasms4]{article}

\newcommand{\go}{\gtrsim}
\newcommand{\lo}{\lesssim}

\begin{document}

\title{
On the Formation and Evolution of Common Envelope Systems
}

\author{Frederic A.\ Rasio}
\affil{Department of Physics, MIT 6--201, Cambridge, MA 02139\\
       Email: rasio@mit.edu}
\and
\author{Mario Livio}
\affil{Space Telescope Science Institute, 3700 San Martin Drive,
  Baltimore, MD 21218\\
  Email: mlivio@stsci.edu}

\begin{abstract}
We discuss the formation of a common envelope system
following dynamically unstable mass transfer in a close binary,
and the subsequent dynamical evolution and final fate of the envelope.
We base our discussion on new three-dimensional hydrodynamic calculations
that we have performed for a close binary system containing
a $4\,M_\odot$ red giant with a $0.7\,M_\odot$ main-sequence star
companion. The initial parameters are chosen to model the
formation of a system resembling V~471~Tau, a typical progenitor
of a cataclysmic variable binary.
The calculations are performed using the smoothed particle
hydrodynamics (SPH) method with up to $5\times10^4$ particles.
As initial condition we use an
exact hydrostatic equilibrium configuration at the onset
of dynamically unstable mass transfer. The nonlinear development of the
instability is followed using SPH until a quasi-static common envelope
configuration is formed. In our highest-resolution calculation, we
find evidence for a corotating region of gas around the central binary.
This is in agreement with the theoretical model proposed by Meyer
\& Meyer-Hofmeister (1979) for the evolution of common envelope
systems, in which this central corotating region is coupled to
the envelope through viscous angular momentum transport only.
We also find evidence that the envelope is convectively unstable,
in which case the viscous dissipation time could be as short as $\sim100$
dynamical times, leading to rapid ejection of the envelope. For V~471~Tau,
our results, and the observed parameters of the system, are entirely
consistent with rapid envelope ejection on a timescale $\sim1\,$yr
and an efficiency parameter $\alpha_{CE}\simeq1$.
\end{abstract}

\keywords{hydrodynamics -- instabilities --
 stars: binaries: close -- stars: evolution -- stars: interiors}

\section{INTRODUCTION AND MOTIVATION}

Common envelope (hereafter CE) evolution is thought to be the
consequence of a dynamical mass transfer event in a close binary system.
A dynamical mass transfer instability can occur when
mass is being transferred from the more massive to the less massive
component and the mass donor has a deep convective envelope,
as in the case of a giant or AGB star losing mass to a less massive
companion. When these conditions are realized,
the mass losing star is usually unable to
contract as rapidly as its Roche lobe and thus it starts
transferring mass on a dynamical timescale.  Typically, under such
conditions, the secondary is unable to accrete all the proffered
mass, and it is driven out of thermal equilibrium.
The system is then
expected to quickly reach a configuration in which the core of the
evolved star and the companion are orbiting each other inside a common
envelope of gas which is not corotating with the binary
(see recent reviews by Webbink 1988; Iben \& Livio 1993; Livio 1996).

The main effect of the CE phase is expected to be a substantial
reduction in the separation of the binary, possibly accompanied by
the ejection of the entire envelope.  The concept of CE evolution, as
described above, was introduced by Paczy\'nski (1976) to explain
the formation of cataclysmic variables and by Ostriker (1975)
in the context of massive X-ray binaries.
It is now thought that all binary systems
containing at least one compact component and with orbital
periods shorter than a few days (this includes most cataclysmic variables,
binary
pulsars, and X-ray binaries) must have gone through a common envelope
phase (except perhaps in dense cluster environments, where close stellar
encounters
can operate). It is not surprising, therefore, that this topic has
received considerable attention in recent years and has been the
subject of many theoretical studies
(de Kool 1987, 1990; Livio \& Soker 1988; Terman et al.\ 1994, 1995 and
references
therein). Most of these theoretical studies have focused on the final stages
of the CE evolution, in an attempt to determine the orbital parameters
of the emerging binary (Taam \& Bodenheimer 1991; Taam et al.\ 1994;
Yorke et al.\ 1995). However, the treatment of the formation and early
dynamical
evolution of CE systems has been extremely crude until now.  For example, a
fictitious drag force was applied to trigger the initial spiral-in in
the calculation of Terman et al.\ (1994)
and the secondary was placed directly inside the envelope of the
primary in Terman et al.\ (1995).

There are several reasons why it is important to study this early phase of CE
evolution.
(i)~This is the only phase in the life of a
CE system that can be modeled in a realistic way. Indeed, current
three-dimensional
hydrodynamic codes can only follow reliably the {\em dynamical phase\/} of the
evolution.
The later stages depend crucially on angular momentum and energy
transport processes that are difficult to model and take place on dissipation
timescales,
which can be many orders of magnitude longer than the dynamical time.
(ii)~It is clear that an accurate calculation of the early dynamical phase is
required
in order to establish the {\em initial conditions\/} of separate calculations
for
the dissipative phase. Attempts to predict the final outcome of a CE
phase based on the efficiency parameter for energy deposition $\alpha_{CE}$
(Livio \& Soker 1988) give results that depend critically on the assumptions
made about these initial configurations. Indeed, because of the
uncertainty in the initial configuration (and the ambiguity in
estimating its binding energy), various  {\em definitions\/} of
$\alpha_{CE}$, e.g.\ in Iben \& Tutukov (1984) and de Kool (1990),
can give results that differ by as much as a factor $\sim 10$
(Yungelson, Tutukov \& Livio 1993).
(iii)~In some cases it is possible that the complete ejection of the envelope
could
occur on a timescale not much longer than the dynamical time (say $\sim100$
dynamical
times). In such a case, a purely dynamical calculation can in fact provide a
complete
description of the CE phase.

In this paper, we present the results of new hydrodynamic calculations in three
dimensions
for the unstable mass transfer in a close binary and
the early dynamical evolution of a CE system. The calculations are done
for a binary containing a red giant with a less massive main-sequence star
companion,
representing a typical progenitor of a pre-cataclysmic-variable system.
The initial conditions for these calculations are carefully constructed
equilibrium configurations for the close binary at the onset of dynamically
unstable mass transfer.
In \S 2 we review our numerical method, smoothed particle hydrodynamics (SPH)
and we describe how to construct equilibrium configurations for close binaries
in three dimensions.
In \S 3 we present and discuss our numerical results.
A summary and conclusions follow in \S4.

\section{NUMERICAL METHOD}

\subsection{The Smoothed Particle Hydrodynamics Code}

The smoothed particle hydrodynamics (SPH) method has been used
for the calculations presented here.
SPH is a Lagrangian method that was introduced
to treat astrophysical problems involving
self-gravitating fluids moving freely in three dimensions
(see Monaghan 1992 for a recent review).
Our SPH code was  developed originally by
Rasio (1991), specifically for the study of close stellar interactions
(Rasio \& Shapiro 1991, 1992). The implementation of the SPH scheme is
similar to that adopted by Hernquist \& Katz (1989), but
the gravitational field is calculated using a fast grid-based FFT solver.
The neighbor searching is performed  using
a multigrid hierarchical version
of the linked-list method usually adopted in P$^3$M particle codes (Hockney \&
Eastwood 1988).
Other details about the implementation, as well as a number of
test-bed calculations using our SPH code for binary systems are presented in
the above references.

The highest-resolution calculation presented here
was done using $N=5\times10^4$ SPH particles,
and each particle interacting with a nearly constant number of neighbors
$N_N\simeq60$. The gravitational potential is calculated by FFT on a $256^3$
grid.
With these resources, a complete calculation, starting from the onset of mass
transfer and ending with the formation of a quasi-static CE configuration,
takes
about 250 CPU hours on an IBM SP-2 supercomputer.

We use a constant number density of SPH particles with varying
particle masses to construct the
initial conditions. This is done in order to maintain good spatial resolution
and mass resolution near the stellar surface, which is particularly
important for problems involving tidal interactions and
 mass transfer in close binaries.
We use a simple ideal gas equation of state with $P=A\rho^{5/3}$,
where $P$ is the pressure, $\rho$ is the density, and $A\propto\exp(s)$ is a
function of
the local specific entropy $s$. Our implementation of SPH
uses $A$ as a fundamental variable and integrates an evolution equation
for entropy rather than energy (Rasio \& Shapiro 1992).
Shocks are the only source of entropy production in
these calculations and we use the standard SPH artificial viscosity
to treat them.

\subsection{Conventions and Choice of Units}

We consider a binary system containing a red giant primary and
a main-sequence secondary. The primary has a mass $M_1=4\,M_\odot$.
The red giant is modeled as a compact core plus an extended envelope, with a
mass ratio
$M_{\rm core}/M_{\rm env}=1/5$, i.e., $M_{\rm core}\simeq0.7\,M_\odot$
and $M_{\rm env}\simeq3.3\,M_\odot$. The mass of the secondary is
$M_2=M_{\rm core}\simeq0.7\,M_\odot$.
Both the secondary and the red-giant core
are modeled as point masses interacting with the gas through gravity
only. Close to the point masses, the gravitational field is smoothed
over a length $h_c$ comparable to the local SPH smoothing length. Specifically,
we treat the core and the secondary as uniform-density spheres of radius $h_c$
when computing their gravitational interaction with an SPH particle.
Typically we have $h_c/R_1\sim10^{-2}$, where $R_1$ is the radius of the
red-giant
envelope. The gravitational interaction between the two point masses themselves
is
not smoothed. Initially
the envelope of the red giant is assumed to have constant specific entropy,
i.e., we
let $A=\,$constant for all SPH particles at $t=0$.

Throughout this paper, numerical results are given in units
where $G=M_{\rm env}=A=1$, where $G$ is the gravitational constant, $M_{\rm
env}$ is the
total mass of {\it gas\/} in the system (initially inside the
red giant envelope), and $A$ is the (constant) entropy variable at $t=0$.
In these units, the binary separation at $t=0$ (onset of mass transfer)
is $r_i\simeq2.9$ and the radius of the red giant is $R_1\simeq1.8$.
The units of time, velocity, and density are then
\begin{eqnarray}
t_o & \simeq &2 \,{\rm d}\times \left(\frac{r_i}{100\,R_\odot}\right)^{3/2}
        \left(\frac{M_{\rm env}}{3.3\,M_\odot}\right)^{-1/2} \\
v_o & \simeq & 140\,{\rm km}\,{\rm s}^{-1} \times
\left(\frac{r_i}{100\,R_\odot}\right)^{-1/2}
       \left(\frac{M_{\rm env}}{3.3\,M_\odot}\right)^{1/2} \\
\rho_o & \simeq & 5\times10^{-4}\,{\rm g}\,{\rm cm}^{-3}
            \times \left(\frac{r_i}{100\,R_\odot}\right)^{-3}
           \left(\frac{M_{\rm env}}{3.3\,M_\odot}\right)
\end{eqnarray}

\subsection{Constructing the Initial Condition}

In addition to its normal use for dynamical calculations, SPH can also
be used to construct highly accurate hydrostatic equilibrium configurations in
three dimensions (Rasio \& Shapiro 1994, 1995).
We consider only {\em synchronized\/} binary configurations in this paper.
This is a reasonable assumption for a system with moderate mass ratio, and we
have
checked that the synchronized configuration we construct at $t=0$ is indeed
tidally stable (although marginally). Binary systems with more extreme mass
ratios
can never reach a stable synchronized configuration and in such a
case an initial condition containing a nonspinning primary may be more
realistic
(see, e.g., Rasio 1996 and references therein).

For a synchronized system, the entire mass of fluid is
in uniform rotation and equilibrium solutions
can be constructed by simply adding a linear friction term
$-{\bf v}/t_{relax}$ to
the Euler equations of motion in the corotating frame.
This forces the fluid to relax to a minimum-energy state.
We use $t_{relax}=1$ in our units, which makes the damping of unwanted
oscillations nearly critical and optimizes the computation time to
converge toward an equilibrium.
Initially, a spherical envelope of constant specific entropy and density is
used for the
red giant and the two stars are placed at a large separation $r\simeq4R_1$.
While the relaxation takes place, the particle entropies are maintained
constant
and their positions are continuously adjusted (by a simple uniform translation
along the binary axis)
so that the separation between the two centers of mass
remains constant. Simultaneously,
the angular velocity $\Omega$ defining the corotating frame is continuously
updated so that the net centrifugal and gravitational accelerations
of the two centers of mass cancel exactly.
With large enough numbers of SPH particles ($N\sim10^4-10^5$), very accurate
equilibrium
solutions can be constructed using this relaxation technique, with the
virial theorem satisfied to an accuracy of about one part in $10^3$ and
excellent agreement found with analytic solutions. In addition, stable
equilibrium
configurations can be maintained accurately during dynamical integrations
lasting
up to $\sim100$ dynamical times (Rasio \& Shapiro 1994, 1995).

In order to construct an initial condition corresponding to the onset of mass
transfer,
the separation $r$ between the centers of mass during the relaxation
calculation is
decreased very slowly (on a time scale much longer than
$t_{relax}$) so that an entire equilibrium {\em sequence\/} is constructed
in a single integration.
In practice we let $r(t)=r(0)-t/t_{scan}$, where $r(0)$ corresponds to
well separated components and $t_{scan}=100$ in our units.
Using this procedure, we can determine precisely the Roche limit configuration
(equilibrium configuration with minimum $r$) simply by observing the
moment when the first few SPH particles begin drifting across the inner
Lagrangian point, onto the secondary (Rasio \& Shapiro 1995). This Roche limit
configuration,
corresponding to the onset of mass transfer, is then used as an initial
condition for a
dynamical integration.

A particle plot corresponding to  the initial Roche limit configuration is
shown in Fig.~1 ($t=0$). The large tidal distortion of the envelope near the
axis is
evident and there are just a few particles that have crossed the inner
Lagrangian point.

%%%%%%%%%%%%%%%%%%%%%%%%%%%%%%%%%%%%%%%%%%%%%%%%%%%%%%%%%%%%%%%%%%%%%%%%%%%

\section{RESULTS}

\subsection{Initial Spiral-In}

Soon after the beginning of mass transfer, a thick torus of material forms
around the
secondary ($t\simeq10-20$). As the initial stream leaving the primary is
compressed rapidly
in a convergent flow and later self-intersects at
a highly supersonic speed, the gas is strongly shocked and the torus expands
rapidly
and thickens, leading to a radial outflow around the secondary in addition to
the
beginning of an accretion flow (cf.\ Sawada et al.\ 1986).
The mass transfer rate increases rapidly, on a timescale
comparable to the orbital period.
Two factors contribute to making the mass transfer dynamically unstable here
(for general discussions of unstable mass transfer, see Hjellming \& Webbink
1987;
Hjellming 1989; Hut \& Paczy\'nski 1984). First, as the red giant
primary loses mass, the isentropic envelope responds adiabatically by
expanding.
Second, since
the mass transfer is from the more massive to the less massive component,
and since angular momentum
is lost from the orbit, the orbital separation tends to decrease. The two
effects
 combine to make
Roche lobe overflow accelerate catastrophically on a dynamical timescale.
By $t\simeq30$, the companion is entering the dense outer layers of the red
giant envelope
and will then plunge in almost radially.

In this dynamical calculation, the spiral-in phase is very brief, taking just
about
one orbital period. This is because we start the calculation when a dynamically
significant
amount of mass is already being transferred. In reality, of course, this phase
is
expected to be much longer, as the Roche lobe eats slowly into the outer
atmosphere of
the primary (e.g., Webbink 1984).

Figure~2 shows the evolution of the separation between the two point masses
during
the entire calculation. The orbital decay is very slow at first, but
accelerates
catastrophically around $t\simeq30$.
As soon as it enters the denser region of gas inside the red giant envelope,
the companion sinks relatively quickly into the deep interior of the envelope,
with
the separation decreasing to about half a stellar radius by $t\simeq32$.
This phase of the evolution is characterized by a much stronger dynamical
interaction
between the companion and the entire mass of gas in the system.

\subsection{Strong Dynamical Interaction}

The motion of the companion inside the envelope is largely supersonic,
especially in the
outer layers where large Mach numbers are reached. As a result, during the
initial phase of the radial plunge, strong dissipation occurs in a simple
bow-shock
structure behind the
companion. This is illustrated in Figure~3, which shows regions of high and low
entropy
in the gas. Here we use $A>2$ to separate the ``high-entropy'' particles.
Recall that all
particles have $A=1$ at $t=0$. In our units, a change of order unity in $A$
over a distance
of order unity corresponds to a buoyancy force comparable to gravity (i.e.,
$dA/dr\sim1$
gives a Brunt-V\"ais\"al\"a frequency comparable to the local dynamical
frequency
$(G\rho)^{-1/2}$).
By $t\simeq35$, however, the red giant core itself has been displaced away from
its original position at the center of the envelope, and the two point masses
are now
truly orbiting one another inside a common gravitational potential well. The
dissipation process becomes much more complex at this point (Fig.~3c,d),
with several interacting
spiral shock fronts propagating out.  These spiral shocks provide the basic
coupling mechanism between the gas and the binary at the beginning of
this strongly dynamical phase of the evolution, while viscous dissipation
becomes
dominant near the end (see below).

It is during this phase of strong dynamical interaction between the two point
masses and the
gas that a large redistribution of energy and angular momentum takes place in
the system.
Figures~4 and~5 show the transfer of energy and angular momentum from the
orbital motion
to the gas during the dynamical evolution.
The energy transfer rate appears approximately constant
during the entire phase of strong dynamical interaction ($t\go30$). Instead,
the orbital
 angular
momentum is lost mostly during the initial spiral-in and the first radial
plunge, but then
becomes approximately constant for $t\go35$. By the end of the calculation,
essentially all
of the angular momentum in the system has been transferred to the gas.
Linear momentum is also exchanged between the gas and point masses and, since
the gas
ejection is not isotropic, the binary receives a small recoil velocity
$v_{rec}$.
At the end of our calculation we find $v_{rec}\simeq0.054$ in our units, which
corresponds to $v_{rec}\simeq8\,{\rm km}\,{\rm s}^{-1}$ for a typical system.

\subsection{The Common Envelope Configuration}

An important question concerns the motion of the gas in the immediate vicinity
of
the binary system after the formation of the common envelope. In our highest
resolution
calculation, {\em we find that
a corotating region of gas forms at the center of the common envelope\/}.
The corotating gas is concentrated near the orbital plane.
This is illustrated in Figure~6, which shows the distribution of SPH particles
in the
$(\Omega, r_{cyl})$ plane at $t=40$.
Here $\Omega=v_t/r_{cyl}$, where $v_t$ is the component of
the particle velocity in the azimuthal direction and $r_{cyl}$ is the
distance to the rotation axis (vertical axis passing through the center of mass
of the binary). Both the equatorial radius and the vertical thickness
of the corotating region are roughly comparable to the binary separation.
This corotating region continues to exist and remains well resolved until the
end
of our highest-resolution calculation (but see \S3.4 below).

At the end of our calculation ($t\simeq62$), most of the mass in
the common envelope has relaxed to a
quasi-static equilibrium. The value of the virial ratio for the bound gas is
$|2T+W+2U|/|W|\simeq0.01$, where $T$ is the kinetic energy, $W$ is the
gravitational
potential energy, and $U$ is the internal energy (this dimensionless virial
ratio
 would be zero in strict hydrostatic equilibrium).
The envelope is rapidly rotating, with $T/|W|\simeq0.11$, not far from the
secular stability
limit at $T/|W|\simeq0.14$ (e.g., Tassoul 1978).
The mass loss fraction can be determined using the method of Rasio \& Shapiro
(1991), based
on the specific binding energy and enthalpy of each individual particle. Using
this method
we find that
about 10\% of the total mass of gas has become unbound by the end of the
calculation
(but see the discussion in \S3.5 below). The final density $\rho$,
specific entropy $A=P/\rho^{5/3}$, and specific angular momentum $j= r_{cyl}
v_t$
inside the common envelope are shown in Figure~7 as a function of the interior
mass
fraction $m/M$. The quantities shown have been mass-averaged over cylindrical
shells
centered on the rotation axis. The correspondence between the interior mass
fraction
$m/M$ and the radius $r_{cyl}$ of a cylinder is also shown (Fig.~7b). The
outermost
$\sim10$\% of the mass is unbound. For the bound gas, the specific angular
momentum
increases with $r_{cyl}$, as required for dynamical stability (e.g., Tassoul
1978).

The specific entropy of the bound gas is nearly constant or increasing
everywhere except
near the center, close to the binary, where it decreases outwards. This is
a clear sign that {\em most of the envelope is or has been dynamically
unstable to convection\/}. The specific
entropy profile left behind by the dynamical interaction between the shrinking
binary and the
envelope has $dA/dr_{cyl}<0$. Convective (Rayleigh-Taylor) instabilities then
develop,
redistributing the material so that a nearly constant specific entropy profile
is obtained.
Close to the binary, however, the convective motions have not yet had the time
to develop
and an unstable entropy gradient is still seen.
Careful examinations (e.g., using computer animations) of the fluid motions
in the inner envelope do indeed reveal the presence of convective motions. On a
dynamical
timescale, and with the fairly coarse spatial resolution of a three-dimensional
calculation, these convective motions take the form of large bubbles of
higher-entropy gas
rising through a region of lower-entropy gas that surrounds them. A clear
example
of such a rising high-entropy bubble (by far the largest observed in this
calculation)
is visible to the left of the binary in the particle plot of Figure~3c (and
Fig.~3d).
Such large inhomogeneities in the gas near the center also produce torques on
the binary,
leading to the somewhat irregular variation of the angular momentum near the
end of the
calculation (Fig.~4, $t>50$).

\subsection{Dependence of the Results on the Numerical Resolution}

Our SPH calculations of the CE evolution must be terminated when the
separation between the two inner cores becomes
comparable to the local spatial resolution in the central region of gas.
Otherwise the interaction between the orbital motion and the gas can no longer
be treated correctly. The conditions at the end of our highest-resolution run
are
illustrated in Figure~8.

In a calculation with a smaller number of particles one may be tempted to
continue past this point. This may be necessary for the gas to reach a
quasi-steady state at the end of the calculation. However, misleading results
can then be obtained. Most importantly for this problem, the motion of the gas
in the
vicinity of the binary system at the end of the calculation will be corrupted
by
the loss of spatial resolution there. If we repeat the same
calculation as above with $N=8000$ particles (which decreases the spatial
resolution
by a factor $\sim2$), and integrate until
the same final binary separation is reached,
we no longer find that the gas is corotating with the
binary at the end. This is shown in Figure~9, where we compare the final
rotation
profiles for the two calculations. The reason is simply that
all SPH particles within a region of size $\sim\langle h\rangle$ close to the
center
(where $\langle h\rangle$ is the average SPH smoothing length in that region)
must necessarily have comparable values of $\Omega$. Therefore, the true
angular
velocity profile near the center
of any rapidly rotating configuration will always be truncated at some distance
$r_{cyl}\sim\langle h\rangle$ inside which $\Omega$ will appear nearly
constant.
In Figure~9, we see that, with a lower-resolution calculation, we would have
concluded incorrectly that the angular velocity of the binary is $\sim10$ times
larger than that of the gas near the center of the final configuration.

Having found that the main coupling mechanism between the binary and the
gas in the final steady state is viscous transport of angular momentum in the
differentially rotating envelope, we must now also
ask how accurately this can be represented in numerical calculations.
SPH calculations in particular can have large spurious shear viscosity because
of
the representation of the fluid by discrete particles, and because all forms of
artificial viscosity (introduced in the method to treat shocks) also introduce
artificial shear viscosity. Extensive tests of SPH using simple experiments
with
shear flows indicate that the angular momentum transport timescale in
differentially
rotating envelopes may be typically $\sim100$ dynamical times
(Lombardi, Rasio, \& Shapiro 1995). This is not much longer than the time span
of
the present calculations, and therefore
we expect the numerical viscosity to play a significant role in our results.

Indeed, if, for example, we repeat short segments of the numerical
integration around $t=50$ with different values for the parameters of the
artificial viscosity, we find that the value of the energy transfer rate (slope
of
the curves near $t=50$ in Fig.~4) changes slightly. If we vary the artificial
viscosity parameters over the full range of values that would still lead to a
reasonable treatment of shocks (smoothing a simple one-dimensional shock front
over a distance $\simeq0.5\langle h\rangle - 5\,\langle h\rangle$),
the energy transfer rate varies by as much as $\sim50$\%.
However, a large effective viscosity may well be present in the real system if,
as our results suggest, the initial entropy profile in the envelope as it is
created
is unstable to convection. Although our three-dimensional calculations cannot
resolve
the fine details of the convective motions, a large eddy viscosity is expected
in this
regime. If processes like convection
(or perhaps magnetic fields; cf.\ Reg\"os \& Tout 1995) determine the effective
viscosity of the envelope in the real system, then we may not be able to
calculate
the evolution in a quantitatively accurate way (a similar problem arises in
models
of accretion disks). In that case, a low-resolution but
qualitatively reasonable calculation such as the one presented here may be our
only
resource for developing a theoretical understanding of CE evolution.

\subsection{The Final Fate of the Envelope}

A naive extrapolation of the energy transfer rate shown in Figure~3 would lead
to the
conclusion that the entire envelope is liberated in a time $t\lo100$, which in
our units
(cf.\ eq.~[2]) means $t\lo1\,$yr. Similarly, if we look at the mass loss
fraction as
a function of time (Fig.~10), a simple extrapolation would predict that all the
gas
becomes unbound after this time. A direct determination of this time is
impossible
here because the calculation must be terminated well before a large fraction of
the mass
has been ejected. In addition, at the end of our calculation, there are large
uncertainties
in the determination of the gas mass fraction that is likely to become unbound
eventually.
This is because we cannot predict how much of the enthalpy associated with each
fluid
particle will be later transformed into kinetic energy of a radial outflow. If
we include
the enthalpy in the estimate of the binding energy (as done in Rasio \& Shapiro
1991),
we obtain an upper limit to the mass loss fraction. If we do not include it we
obtain a
lower limit. The two estimates differ significantly at the end of our
calculation
(Fig.~10).

If the evolution of the CE system indeed continues on this short viscous
transport
time, and if we assume that the binary orbit ceases to evolve when the total
gas energy
becomes positive, then we can predict the final binary separation simply from
conservation of energy (Fig.~4). This gives a final separation $r_f\simeq0.04$
in our
units, or $r_f/r_i\simeq1.5\times10^{-2}$. This number is in reasonable
agreement with
the parameters of V471 Tau (which our initial condition is meant to represent),
where $r_f\simeq3\,R_\odot$ and $r_i\sim100\,R_\odot$ (Iben \& Livio 1993).
Over such a short time, energy transport processes
and radiative losses in the gas are completely negligible, implying that a
proper
definition of the parameter $\alpha_{CE}$ should give $\alpha_{CE}\simeq1$ in
this case.
For comparison, one of the commonly used approximate definitions
\begin{equation}
\frac{(M_1+M_2)M_{env}}{2r_i}\simeq \alpha_{CE}\, M_{core}M_2
  \left(\frac{1}{2r_f} - \frac{1}{2r_i}\right)
\end{equation}
(Iben \& Livio 1993, eq.~[17]) gives $\alpha_{CE}\simeq0.5$ if we insert
$r_f/r_i=1.5\times10^{-2}$ and our values of the masses.

\section{CONCLUSIONS}

The calculations presented in this paper cover the {\em initial
dynamical phase\/} of CE evolution, i.e., a brief episode of
unstable mass transfer followed by rapid spiral-in of the companion
into the primary and the formation of a CE configuration.
For the first time, this dynamical phase of CE evolution has been
followed starting from a realistic initial condition.
Specifically, we have demonstrated that hydrodynamic
calculations can be started from an exact hydrostatic equilibrium solution,
representing a close binary system at the onset of unstable mass transfer,
and can then follow the evolution
 through the entire dynamical phase until a quasi-static
CE configuration is formed.

Our hydrodynamic calculations can treat only a relatively
short evolutionary timescale ($\lo100$ dynamical times, i.e., $\lo1\,$yr for a
typical system). Consequently, the final CE
configuration obtained in the present study could be regarded merely
as the initial condition for a potentially slower spiral-in
phase, which is regulated by viscous dissipation.
The total amount of mass that became unbound during the dynamical phase is
only about 10\% of the envelope mass.
This is to be expected given the short duration of this phase.
We have pointed out, however, that an extrapolation of the energy transfer rate
obtained in the final stages of our calculation could imply the ejection
of the entire envelope on a very short timescale ($\sim1\,$yr) and a high
efficiency
($\alpha_{CE}\simeq1$) of the process. The corresponding reduction in the
binary
separation would be by a factor $r_i/r_f\sim100$, which is sufficient to
explain the
formation of a system like V~471~Tau.

Perhaps our most significant new result is that, during the dynamical phase of
CE evolution,
a corotating region of gas is established near the central binary. This is done
through a combination of spiral shock waves and gravitational torques that can
transfer angular momentum from the binary orbit to the gas, and are most
effective
in the region close to the binary.
The corotating region has the shape of an oblate spheroid encasing the binary
(i.e., the corotating gas is concentrated in the orbital plane).
Our results suggest that the subsequent evolution of the system will be
determined
by the {\em viscous coupling between this rigidly rotating inner
core and the outer, differentially rotating envelope\/}.
The assumption that rigid rotation will be tidally enforced in a core
surrounding the inner binary was already made in the pioneering paper
by Meyer \& Meyer-Hofmeister (1979). Later, clear indications for such a
configuration
were obtained with the first two-dimensional numerical calculations of CE
evolution
(e.g., Sawada et al.\ 1984).
Although the results of some more recent calculations done in three dimensions
(Terman et al.\ 1994, 1995)
appear to indicate that the angular velocity of the gas near the center falls
short
of corotation, we have pointed out that this may well be the result of
insufficient
spatial resolution in those calculations.

Accurate calculations of the dissipative phase of CE evolution, including
viscous transport of angular momentum and energy transport in the envelope,
will be very difficult, given the large uncertainties in our basic
understanding of these
processes in general. If, however, as our results suggest, the envelope is
convective,
the effective viscous dissipation rate may be large enough to eject the entire
envelope on a timescale as short as $\sim100$ dynamical times.
In that case, a simple extrapolation of our dynamical results, as mentioned
above,
may in fact provide a reasonably accurate description of the entire CE
evolution.

\acknowledgments

We thank the Institute of Astronomy at the University of Cambridge, UK,
for hospitality while this paper was being written.
This work was begun at the Institute for Advanced Study in Princeton,
where F.A.R.\ was supported by a Hubble Fellowship, funded by NASA
through Grant HF-1037.01-92A from the Space Telescope Science Institute,
which is operated by AURA, Inc., for NASA, under contract NAS5-26555.
M.L.\ acknowledges support from NASA Grant NAGW-2678.
Computations were performed on the IBM SP-2 parallel supercomputer
at the Cornell Theory Center,
which receives major funding from the NSF and from IBM Corporation,
with additional support from the New York State Science and Technology
Foundation and members of the Corporate Research Institute.

\clearpage

%FIG1
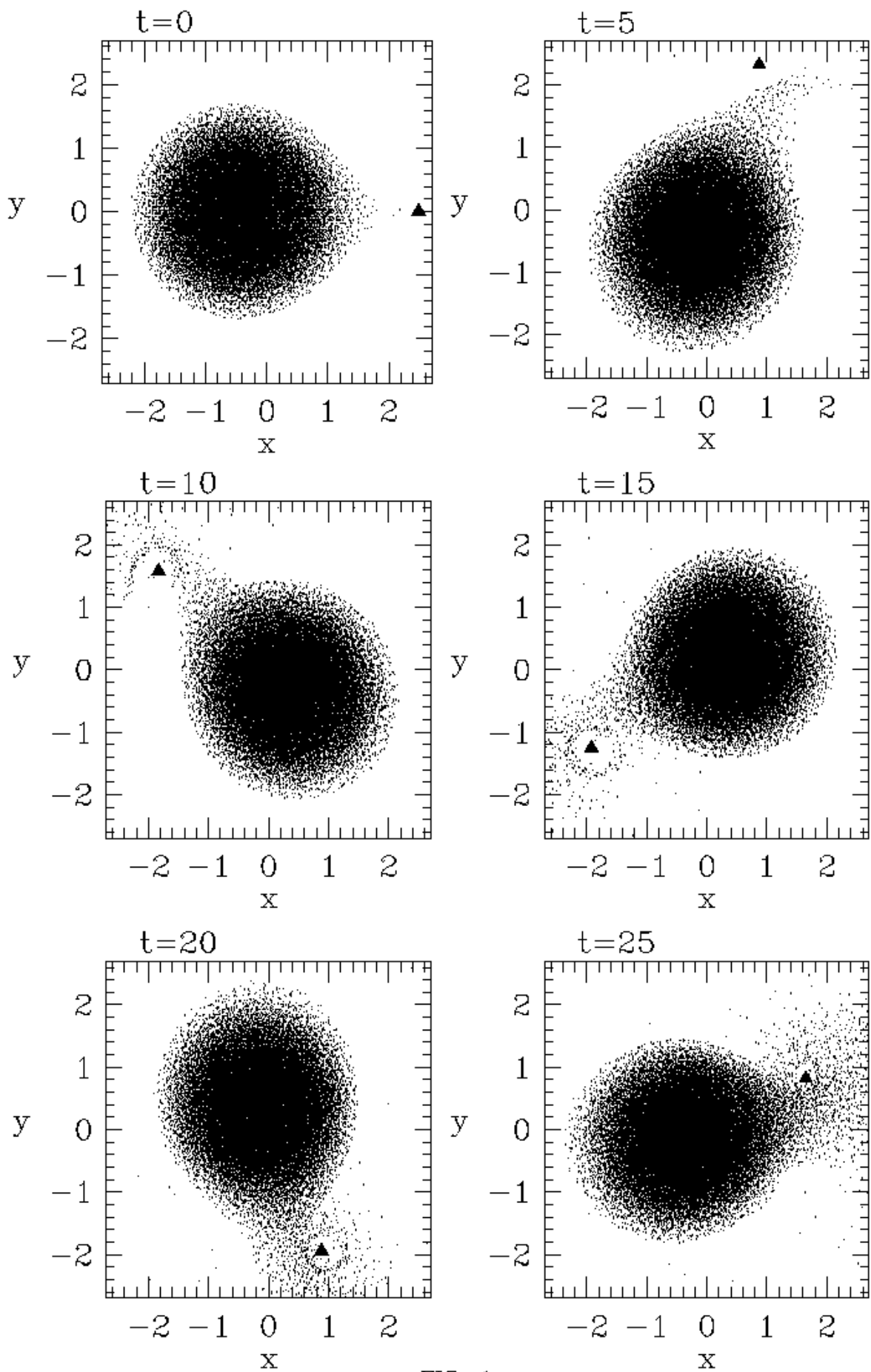
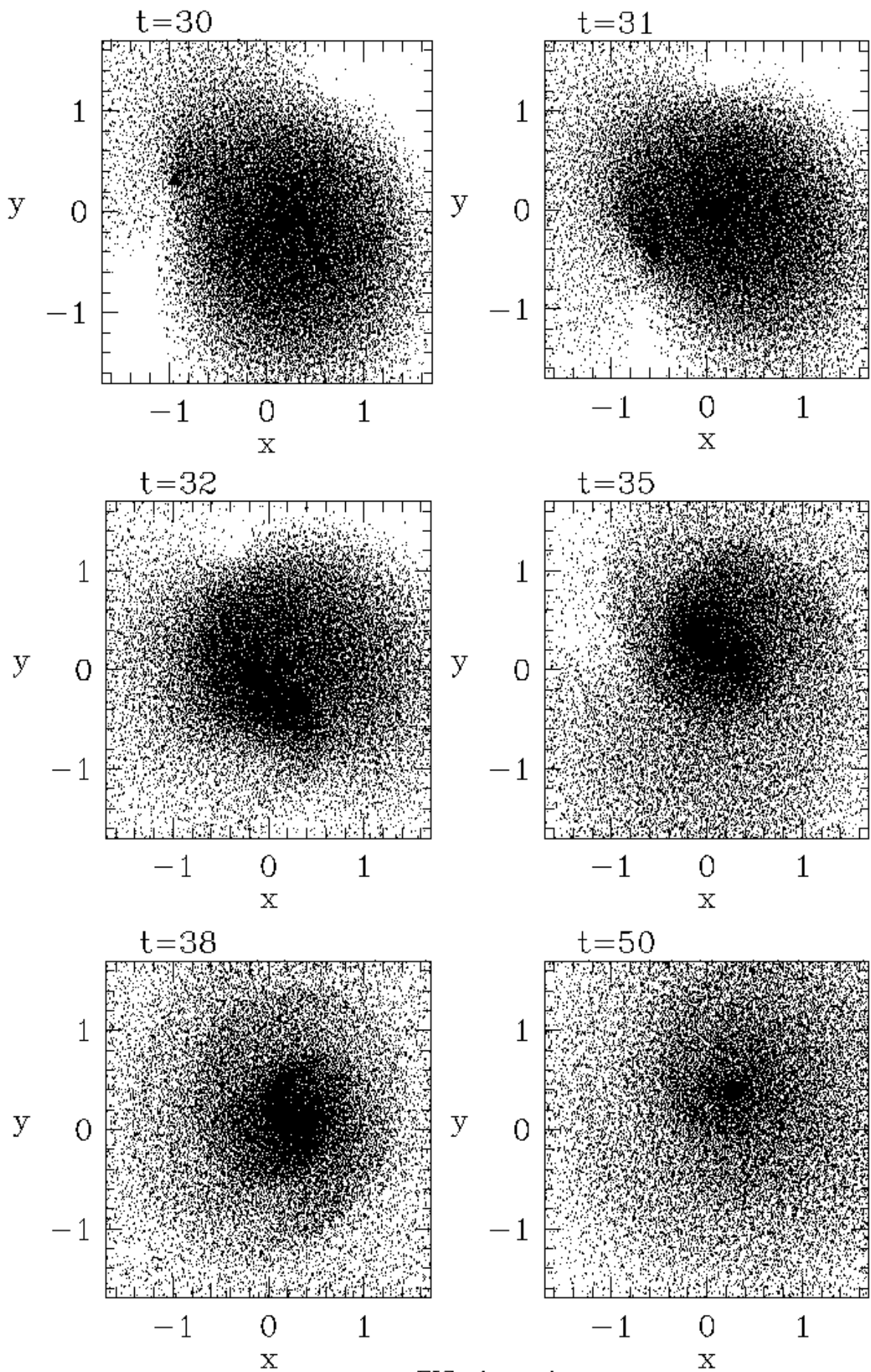
\begin{figure}
\caption{
Particle plots showing the evolution of the system from the onset of unstable
mass transfer
to the formation of the common envelope. Projections of all SPH particles onto
the orbital
$(x,y)$ plane of the binary are shown at various times. The origin is at the
center of mass
of the system. Units are defined in \S2.2.
The core of the red giant is marked by a
large round dot, the companion by a triangle. Both are treated as point
masses in the calculation, but their gravitational interaction with the gas is
softened
over a distance roughly comparable to the size of the symbols in these plots.
The orbital rotation is counterclockwise.
At $t=0$ the system is in a synchronized hydrostatic
equilibrium configuration with just a few SPH particles outside the critical
Roche lobe of
the primary. By $t\simeq20$ the mass transfer flow is well established and the
orbit is decaying
unstably. From $t\simeq30$ to $t\simeq50$ the companion quickly spirals in
through the
envelope and the common envelope system is formed. The calculation ends when
the separation between the two
point masses becomes comparable to the local SPH smoothing length (also
comparable to the
gravitational smoothing length).
}
\end{figure}

%FIG2
\begin{figure}
\caption{
Time evolution of the separation between the two point masses in the
calculation.
In the insert we show the details of the evolution near the end of the
calculation,
on a linear scale. The separation is given here in units of the initial binary
separation $r_i$, while time is in the units defined in \S2.2.
}
\end{figure}

%FIG3
\begin{figure}
\caption{
Particle plots showing regions of high and low entropy in the gas near the
equatorial
plane ($|z|<0.1$). All fluid particles
have $A=P/\rho^{5/3}=1$ at $t=0$. The left and right plots show particles with
$A>2$
(high entropy) and $A<2$ (low entropy), respectively.
In~(a) and~(b), we show the simple bow-shock structure
behind the companion in the early phase of the radial plunge (here for $t=32$;
see the
corresponding plot in Fig.~1). In~(c) and~(d), a more complex structure is seen
($t=50$),
following the displacement of the red giant core and the interactions between
various spiral
shock fronts.
}
\end{figure}

%FIG4
\begin{figure}
\caption{
Time evolution of various energies in the system. The orbital energy is that
associated
with the two point masses only (note that this is not equal to the orbital
energy of the
binary system at $t=0$ since the primary also contains all the gas at that
time). The
gas energy is the sum of the kinetic, thermal, and self-gravitational energies
associated
with all SPH particles, plus the potential energy of the (softened)
gravitational interaction between
the SPH particles and the two point masses. The total (conserved) energy is
then the sum
of the gas and orbital energies.
}
\end{figure}

%FIG5
\begin{figure}
\caption{
Time evolution of the angular momentum in the system. The orbital angular
momentum is
that associated with the motion of the two point masses only, while the gas
angular
momentum corresponds to the motion of all SPH particles.
}
\end{figure}

%FIG6
\begin{figure}
\caption{
Angular velocity profiles in the common envelope at $t=40$. The
large dot indicates the position of the two orbiting point masses.
The lighter dots show the positions of individual SPH particles.
The three plots correspond to horizontal slices (perpendicular to the rotation
axis)
progressively further away from the equatorial (symmetry) plane $z=0$.
Clearly, a corotating region of gas exists close to the binary system near
$z=0$.
}
\end{figure}

%FIG7
\begin{figure}
\caption{
Average density ($\rho$), specific entropy ($A= P/\rho^{5/3}$), and specific
angular momentum ($j= r_{cyl} v_t$) profiles in the
common envelope at the end of the calculation. Here $m/M$ is the gas mass
fraction
interior to a cylindrical surface centered on the rotation axis
($m/M=0$ at the center and $m/M=1$ at the outer surface) and
$r_{cyl}$ is the distance to the rotation axis.
}
\end{figure}

%FIG8
\begin{figure}
\caption{
Particle plot showing the region of gas close to the binary system at the end
of the calculation. The two large circles indicate the extent of the regions
where the gravitational interaction between the point masses and the gas is
smoothed.
This is comparable to the average SPH smoothing length $\langle h\rangle$
in this region, also shown.
}
\end{figure}

%FIG9
\begin{figure}
\caption{
Angular velocity profile in the the common envelope at the end of two
calculations
with different spatial resolution. The square dots correspond to our
highest-resolution
calculation, with $N=5\times10^4$ particles, whereas the triangles correspond
to a
calculation with $N=8000$. The central corotating region of gas is no longer
resolved.
}
\end{figure}

%FIG10
\begin{figure}
\caption{
Mass loss fraction $\Delta M_{loss}/M$ at various times during the strong
dynamical
interaction phase and at the end of the calculation. The triangles give a lower
limit, based on an estimate of the binding energy of each fluid particle that
does
not include its enthalpy. The squares give an upper limit, assuming that all
the
enthalpy is eventually transformed into kinetic energy of an outflow.
}
\end{figure}

\end{document}